\documentclass[letterpaper, 10 pt, conference]{ieeeconf}  

\IEEEoverridecommandlockouts                              

\title{\LARGE \bf
The Silver Ratio and its Relation to Controllability
}

\author{Michael Muehlebach$^{1}$
\thanks{$^{1}$Michael Muehlebach is with RISELab at the University of California, Berkeley, California, USA. E-mail: {\tt\small michaelm@berkeley.edu}.}%
}

\usepackage[english,american]{babel}
\usepackage{url}
\usepackage{epsfig} 
\usepackage{times} 
\usepackage{amsmath}
\usepackage{amsfonts}
\usepackage{amssymb}  

\usepackage{xcolor}
\usepackage{transparent}
\usepackage{graphicx}
\graphicspath{{media/}}
\usepackage{tikz}
\usepackage{epstopdf}
\usepackage{units}
\usepackage{pgfplots}
\usepackage{algorithm}
\usepackage{algpseudocode}
\usepackage{ifthen}

\usepackage{pstool}

\graphicspath{{media/}}
\newcommand{\executeiffilenewer}[3]{%
 \ifnum\pdfstrcmp{\pdffilemoddate{#1}}%
 {\pdffilemoddate{#2}}>0%
 {\immediate\write18{#3}}\fi%
}

\usepackage{arydshln}
\usepackage{perpage}
\MakePerPage{footnote}

\pgfplotsset{every tick label/.append style={font=\footnotesize}}

\newcommand{\diag}{\mathop{\mathrm{diag}}}

\newcommand{\TT}{\ensuremath{\mathsf{\tiny{T}}}}
\newcommand{\T}{^{\TT}}

\newcommand{\V}[2][]{{^{\textsc{\scriptsize{#2}}}{#1}}}

\begin{document}

\maketitle
\thispagestyle{empty}
\pagestyle{empty}

\begin{abstract}
This note investigates the controllability of two unstable second-order systems that are coupled through a common input. These dynamics occur for different types of inverted-pendulum systems. Controllability is quantified by the volume of the state-space that can be reached with unit energy, provided that the system starts and ends at the origin. It is shown that controllability is maximized when the ratio between the time constants amounts to the silver ratio.
\end{abstract}

\begin{keywords}
Controllability, number theory, silver ratio, inverted pendulum, control system design
\end{keywords}

\section{Introduction}
When designing mechatronic systems it is often desirable to optimize the mechanical design such that the resulting control problem is facilitated. This not only helps the control engineer to design a stabilizing controller, but also improves the performance of the resulting system. There are several ways in which this can be achieved. For example, one could directly optimize the closed-loop performance resulting from an optimal control design (e.g. $\mathcal{H}_\infty$ or $\mathcal{H}_2$). While this is a very viable approach in practice, it depends on the specific sensors and actuators characteristics, might provide little insights, and might not capture the transient behavior (it is often assumed that the system starts at the origin). In the following, we use a notion of controllability to study the behavior of two unstable second-order systems that are coupled through a common input. More precisely, we quantify controllability as the region of the state-space that can be reached with unit energy, provided that the system's trajectories start and end at the origin. All results are derived in closed form by appropriate reasoning and standard linear algebra. They do, however, depend on the choice of coordinates. We show that controllability is maximized when the ratio between the time constants equals the silver ratio (in a canonical set of coordinates). The second-order dynamics that are studied in the following occur for different inverted-pendulum type systems, including reaction-wheel based pendula, as well as the inverted-pendulum-on-a-cart system with two parallel pendula, \cite{barton}.

The notion of controllability that is used in the following is explained and motivated in \cite{zhou}. There are many articles that establish connections between number theory and optimal control or optimal filtering. In \cite{fibonacci}, the authors show that the Kalman filter's feedback gain of a scalar random walk system with unit process and measurement noise variance is related to the Fibonacci sequence. Moreover, the filter's steady-state variance estimate equals the golden ratio. The results are then further extended for a more general type of system in \cite{extended}. The article \cite{optimalControl} establishes a link between a class optimal control problems for scalar systems and the Fibonacci sequence, whereas \cite{donoghue} studies control and estimation problems related to the Fibonacci system (a dynamical system whose impulse response yields the Fibonacci sequence). The authors of \cite{feedbackControlPrime} study a differential equation that models the density of prime numbers and investigate the stability of its solution under perturbations. The golden ratio is also commonly used in optimization, for example in the context of line search, see e.g. \cite{wilde}. 

In this work we establish a connection between the silver ratio and controllability. We also motivate the dynamics that are studied using a canonical inverted pendulum system. In fact, this note arose from the controllability analysis of reaction-wheel inverted pendulum systems such as \cite{cubli}. The resulting roll and pitch dynamics are of the type as discussed below. In particular, the analysis concludes that a single reaction wheel is enough for stable balancing, as long as the roll and pitch axis have different inertia. 
%

\emph{Notation:} Real numbers are denoted by $\mathbb{R}$ and vectors are expressed as $n$-tuples $(x_1,x_2,\dots,x_n)$ with dimension and stacking clear from context. The set of square integrable real-valued functions is denoted by $\mathcal{L}_2$, with $||\cdot||$ the standard $\mathcal{L}_2$-norm. The Euclidean norm is referred to as $|\cdot|$.

\section{Problem Formulation}
We consider the system governed by 
\begin{align}
\ddot{x}_1(t)&=\pi_1^2 x_1(t) + v_1 \pi_1^2 u(t) \label{eq:sys1}\\
\ddot{x}_2(t)&=\pi_2^2 x_2(t) + v_2 \pi_2^2 u(t), \label{eq:sys2}
\end{align}
where $\pi_1$, $\pi_2$, $v_1$, and $v_2$ are real constants and $u(t)$ denotes the control input. In the following, the system' s state is denoted by $z(t):=(x_1(t), \dot{x}_1(t), x_2(t),\dot{x}_2(t))$. The aim is to find the ratio between $\pi_1$ and $\pi_2$ such that the volume of  
\begin{multline}
X:=\bigcup_{||u||^2 \leq 1}\{ z(0)\in \mathbb{R}^4~|~z~\text{satisfies \eqref{eq:sys1} and \eqref{eq:sys2}}, \\[-12pt]
\lim_{t \rightarrow \infty} z(t) = \lim_{t\rightarrow -\infty} z(t) = 0\} \label{eq:defX}
\end{multline} 
is maximized.
The set $X$ describes the region of the state space that can be reached with unit energy, provided that the system starts at the origin and ends at the origin.\footnote{Note that the evaluation $z(0)$ at time $t=0$ is arbitrary, as the dynamics are time-invariant. The set $X$ is thus equal to the union of all states that can possibly be reached, provided the system starts and returns to the origin.} This definition generalizes the classical reachability concept (i.e. the set of states that can be reached with unit energy starting from the origin) to unstable systems, \cite{zhou}.

\section{Motivation}\label{Sec:Motiv}
This section motivates why maximizing $X$ leads to a trade-off between the two time constants $\pi_1$ and $\pi_2$. To that extent, the following auxiliary system 
\begin{align}
\dot{\xi}_1(t)&=-\alpha_1 \xi_1(t) + \beta_1 u(t) \label{eq:sub1}\\
\dot{\xi}_2(t)&=-\alpha_2 \xi_2(t) + \beta_2 u(t), \label{eq:sub2}
\end{align}
is analyzed, where $\alpha_1\neq 0$, $\alpha_2\neq 0$, $\beta_1\neq 0$ and $\beta_2\neq 0$ are real-valued constants. The auxiliary system will describe either the stable or the unstable subspace of the dynamics given by \eqref{eq:sys1} and \eqref{eq:sys2}, and motivates the definition \eqref{eq:defX}.

Provided that $\alpha_1$ and $\alpha_2$ are positive, the set of all states that can be reached (in infinite time) with unit energy starting from the origin is given by
\begin{equation}
X_\text{f}:=\{\xi_\text{f} \in \mathbb{R}^2~|~\xi_\text{f}\T \left(\begin{array}{cc} \frac{\beta_1^2}{2 \alpha_1} &\frac{\beta_1 \beta_2}{\alpha_1+\alpha_2} \\ \frac{\beta_1 \beta_2}{\alpha_1+\alpha_2} & \frac{\beta_2^2}{2\alpha_2} \end{array}\right)^{-1} \xi_\text{f} \leq 1\}.\label{eq:defXf}
\end{equation}
This is a standard result, see for example \cite[Ch.8]{Desoer}.
As long as $\alpha_1 \neq \alpha_2$ and $\beta_1\neq 0$, $\beta_2\neq 0$ this amounts to an ellipse, centered at the origin, whose area is
\begin{equation}
\frac{\beta_1 \beta_2}{2 \sqrt{\alpha_1 \alpha_2}}~ \frac{|\alpha_1 - \alpha_2|}{\alpha_1+\alpha_2}. \label{eq:det1}
\end{equation} 
This formula can be decomposed into three parts: $\beta_1/\sqrt{2\alpha_1}$ describes the region that can be reached when considering \eqref{eq:sub1} only, $\beta_2/\sqrt{2\alpha_2}$ is related to the region that can be reached by \eqref{eq:sub2} only, and $|\alpha_1-\alpha_2|/(\alpha_1+\alpha_2)$ results from the coupling due to the common input. The coupling term depends only on the ratio between $\alpha_1$ and $\alpha_2$. Thus, the area increases for smaller $\alpha_1$ and $\alpha_2$, when leaving $\beta_1$, $\beta_2$, and the ratio $\alpha_1/\alpha_2$ fixed.

It is clear that once the set $X_\text{f}$ is reached (starting from $\lim_{t\rightarrow -\infty}\xi(t)=0$), applying $u=0$ ensures that the system converges to the origin again (we have $\alpha_1 > 0, \alpha_2>0$). We can therefore rewrite the set $X_\text{f}$ as
\begin{multline}
\bigcup_{||u||^2\leq 1}\{ \xi(0) \in \mathbb{R}^2~|~\xi~\text{satisfies \eqref{eq:sub1} and \eqref{eq:sub2}}, \\[-12pt]
\lim_{t\rightarrow -\infty} \xi(t)=\lim_{t\rightarrow \infty}\xi(t)=0 \}.\label{eq:tmp1}
\end{multline}
The above definition encompasses also the case $\alpha_1<0$ and $\alpha_2<0$. In fact, the case $\alpha_1<0$ and $\alpha_2<0$ can be identified with the case $\alpha_1>0$ and $\alpha_2>0$ by running the dynamics backwards in time, which concludes that \eqref{eq:tmp1} is equal to $X_\text{f}$ also in case $\alpha_1<0$ and $\alpha_2<0$.

In case $\alpha_1 <0$, $\alpha_2 >0$, \eqref{eq:tmp1} reduces to
\begin{equation}
\{(\xi_1,\xi_2)\in \mathbb{R}^2~|~(2 |\alpha_1|/\beta_1^2) \xi_1^2+(2 \alpha_2/\beta_2^2) \xi_2^2 \leq 1\}, \label{eq:tmp2}
\end{equation}
as the stable and unstable parts decouple. Indeed, the unstable part limits the set of initial conditions that can be \emph{driven to zero} with unit energy, whereas the stable part limits the set of initial conditions that can be \emph{reached from zero} with unit energy. More precisely, $(2  |\alpha_1|/\beta_2^2) \xi_2^2$ energy units are at least needed to drive \eqref{eq:sub1} and \eqref{eq:sub2} from $(*,\xi_2)$ to the origin and $(2 \alpha_2/\beta_1^2) \xi_1^2$ energy units are at least required to reach $(\xi_1,*)$ from the origin. As a result, the set \eqref{eq:tmp2} is obtained by the superposition of the stable and unstable parts. Its area is larger than \eqref{eq:defXf}. We refer the reader to \cite{zhou} for a rigorous argument.

This indicates that $X$ can be decomposed into the superposition of the stable and unstable subspaces of \eqref{eq:sys1} and \eqref{eq:sys2}. The dynamics on both of these subspaces are of the form \eqref{eq:sub1} and \eqref{eq:sub2} with either $\alpha_1=\pi_1$, $\alpha_2=\pi_2$ and $\alpha_1=-\pi_1$, $\alpha_2=-\pi_2$. The trade-off between the time constants $\pi_1$ and $\pi_2$ thus results from the functional dependence of $\beta_1$ and $\beta_2$ on $\pi_1$ and $\pi_2$.

\section{The silver ratio}\label{Sec:SR}
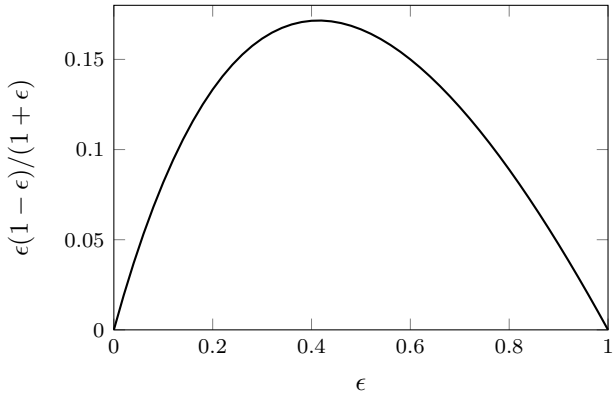
\begin{figure}
\newlength{\figurewidth}
\newlength{\figureheight}
\setlength{\figurewidth}{0.8\columnwidth}
\setlength{\figureheight}{0.5\columnwidth}
%
%
\definecolor{mycolor1}{rgb}{0.00000,0.44700,0.74100}%
\begin{tikzpicture}

\begin{axis}[%
width=0.95092\figurewidth,
height=\figureheight,
at={(0\figurewidth,0\figureheight)},
scale only axis,
xmin=0,
xmax=1,
xlabel={$\epsilon$},
ymin=0,
ymax=0.18,
ylabel={$\epsilon (1-\epsilon)/(1+\epsilon)$},
tick label style={/pgf/number format/fixed},
legend style={legend cell align=left,align=left,draw=white!15!black}
]
\addplot [color=black,solid,thick,forget plot]
  table[row sep=crcr]{%
0	0\\
0.02	0.0192156862745098\\
0.04	0.0369230769230769\\
0.06	0.0532075471698113\\
0.08	0.0681481481481481\\
0.1	0.0818181818181818\\
0.12	0.0942857142857143\\
0.14	0.105614035087719\\
0.16	0.115862068965517\\
0.18	0.125084745762712\\
0.2	0.133333333333333\\
0.22	0.140655737704918\\
0.24	0.147096774193548\\
0.26	0.152698412698413\\
0.28	0.1575\\
0.3	0.161538461538462\\
0.32	0.164848484848485\\
0.34	0.167462686567164\\
0.36	0.169411764705882\\
0.38	0.170724637681159\\
0.4	0.171428571428571\\
0.42	0.171549295774648\\
0.44	0.171111111111111\\
0.46	0.17013698630137\\
0.48	0.168648648648649\\
0.5	0.166666666666667\\
0.52	0.164210526315789\\
0.54	0.161298701298701\\
0.56	0.157948717948718\\
0.58	0.154177215189873\\
0.6	0.15\\
0.62	0.145432098765432\\
0.64	0.140487804878049\\
0.66	0.135180722891566\\
0.68	0.12952380952381\\
0.7	0.123529411764706\\
0.72	0.117209302325581\\
0.74	0.110574712643678\\
0.76	0.103636363636364\\
0.78	0.0964044943820225\\
0.8	0.0888888888888889\\
0.82	0.0810989010989011\\
0.84	0.0730434782608696\\
0.86	0.0647311827956989\\
0.88	0.0561702127659574\\
0.9	0.0473684210526316\\
0.92	0.0383333333333333\\
0.94	0.0290721649484536\\
0.96	0.0195918367346939\\
0.98	0.00989898989898991\\
1	0\\
};
\end{axis}
\end{tikzpicture}%
\caption{Shown is the function $\epsilon (1-\epsilon)/(1+\epsilon)$ for $\epsilon\in (0,1]$, where $\epsilon=\pi_1/\pi_2$ is the ratio between the two time constants. The function is concave for $\epsilon\in (0,1]$ and attains its maximum at $1/\delta_s$.}
\label{Fig:fun}
\end{figure}

We now turn to maximizing the volume of $X$, where $X$ is defined in \eqref{eq:defX}. To simplify the derivation, it is assumed for the moment that $v_1=v_2=1$. For notational convenience we introduce the variables $z_1(t):=(x_1(t),\dot{x}_1(t))$, $z_2(t):=(x_2(t),\dot{x}_2(t))$. Defining the state transformation $z_i(t)=T_i \hat{z}_i(t)$, $i=1,2$, as
\begin{equation}
T_i=\frac{1}{\sqrt{1+\pi_i^2}} \left( \begin{array}{cc} 1 &1  \\ \pi_i &-\pi_i \end{array}\right),
\end{equation}
reveals the dynamics of the stable and unstable subspaces, i.e. for $i=1,2$,
\begin{equation}
\dot{\hat{z}}_i(t)=\left( \begin{array}{cc} \pi_i &0 \\ 0 &-\pi_i\end{array} \right)\hat{z}_i(t)+\frac{\sqrt{1+\pi_i^2}}{2}\left( \begin{array}{c} \pi_i \\ -\pi_i \end{array}\right)u(t).
\end{equation}
These are of the form \eqref{eq:sub1} and \eqref{eq:sub2}. We thus obtain from \eqref{eq:defXf} and \eqref{eq:tmp2} that 
\begin{multline}
X=\{ T \eta \in \mathbb{R}^4~|~(\eta_1,\eta_3)\T P^{-1} (\eta_1,\eta_3) \\
+ (\eta_2,\eta_4)\T P^{-1} (\eta_2,\eta_4) \leq 1\},
\end{multline}
with
\begin{equation}
P:=\frac{1}{4} \left( \begin{array}{cc} \frac{\pi_1}{2} (1+\pi_1^2) & \frac{\pi_1 \pi_2 \sqrt{1+\pi_1^2} \sqrt{1+\pi_2^2}}{\pi_1 + \pi_2} \\
\frac{\pi_1 \pi_2 \sqrt{1+\pi_1^2} \sqrt{1+\pi_2^2}}{\pi_1 + \pi_2} & \frac{\pi_2}{2}(1+\pi_2^2)\end{array}\right)
\end{equation}
and $T:=\diag(T_1,T_2)$.
As a result, the volume of $X$ is proportional to
\begin{align*}
\text{vol}(X)\sim&\left(\det(T^{-\mathsf{\tiny{T}}} \Pi\T \diag(P^{-1},P^{-1}) \Pi  T^{-1})\right)^{-\frac{1}{2}}\\
&=\det(P) \det(T)=\det(P) \det(T_1) \det(T_2),
\end{align*}
where $\Pi$ is a permutation matrix (and therefore has $\det(\Pi)=\pm 1$). This yields
\begin{equation}
\text{vol}(X)\sim\left(\frac{ \pi_1\pi_2}{4} \frac{(\pi_1-\pi_2)}{(\pi_1 + \pi_2)}\right)^2.
\end{equation}
By assuming, without loss of generality, that $0<\pi_1\leq\pi_2$ and by defining the ratio between $\pi_1$ and $\pi_2$ as $\epsilon:=\pi_1/\pi_2$, we obtain
\begin{equation}
\text{vol}(X)\sim \left(\frac{\pi_2^2}{4}~\epsilon \frac{(1-\epsilon)}{(1+\epsilon)}\right)^2,
\end{equation}
with $\epsilon \in (0,1]$, or, in case $v_1\neq 1$ and $v_2 \neq 1$,
\begin{equation}
\text{vol}(X)\sim v_1^2 v_2^2 \left(\frac{\pi_2^2}{4}~\epsilon \frac{(1-\epsilon)}{(1+\epsilon)}\right)^2.
\end{equation}
Thus, the function $\epsilon (1-\epsilon)/(1+\epsilon)$, as plotted in Fig.~\ref{Fig:fun}, describes how the volume of $X$ depends on the ratio between $\pi_1$ and $\pi_2$. The function is concave for $\epsilon\in (0,1]$, and attains its maximum at 
\begin{equation}
\epsilon^*:=\sqrt{2}-1=\frac{1}{\delta_\text{s}},
\end{equation}
which is the inverse of the silver ratio $\delta_\text{s}$. Thus, the volume of $X$ is maximized by\footnote{Assuming that $v_1$ and $v_2$ do not depend on $\pi_1$ and $\pi_2$.} 
\begin{equation}
\pi_2/\pi_1=\delta_\text{s}=1+\sqrt{2}.\label{eq:final}
\end{equation}

\section{Examples}
\begin{figure}
\def\svgwidth{.9\columnwidth}
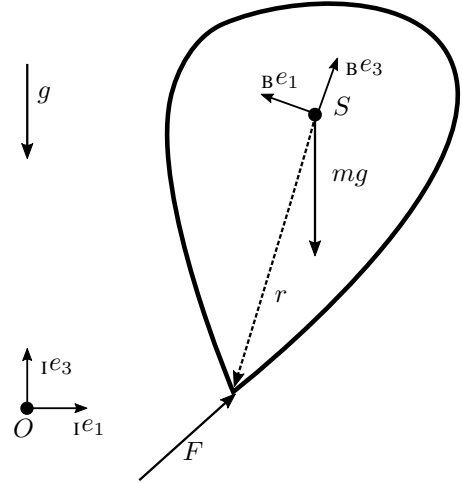
\caption{The figure shows a sketch of the rigid body actuated by the force $F$. The body-fixed frame is indicated by the two unit vectors $_{\textsc{B}}e_1$ and $_{\textsc{B}}e_3$. The two unit vectors $_{\textsc{I}}e_1$ and $_{\textsc{I}}e_3$ belong to the inertial frame of reference. The point $O$ refers to the origin of the inertial frame of reference, whereas the point $S$ represents the center of gravity.}
\label{Fig:RBSketch}
\end{figure}

This section provides a canonical example leading to the dynamics \eqref{eq:sys1} and \eqref{eq:sys2}. We consider an inverted-pendulum system consisting of a rigid body that is balanced about one of its principle axis of inertia. The rigid body is actuated by a single force as shown in Fig.~\ref{Fig:RBSketch}. The force thus acts at a point that is located at the distance $l$ along a principle axis of inertia from the center of gravity. With an appropriate choice of coordinate frame, we obtain the following attitude dynamics
\begin{equation}
\V[\Theta]{B} \V[\dot{\omega}]{B} + \V[\omega]{B} \times \V[\Theta ]{B} \V[\omega]{B} = \V[r]{B} \times \V[F]{B},
\end{equation}
where $\V[\omega]{B}$ denotes the angular velocity represented in the body-fixed frame, $\V[\Theta]{B}=\diag(I_1,I_2,I_3)$ the inertia with respect to the center of gravity represented in the body-fixed frame, and $\V[r]{B}=(0,0,-l)$ the vector from the center of gravity to the point where the force $F$ acts (also expressed in the body-fixed frame). The task will be to control the body's position and attitude about its ``upright" equilibrium. As a result, at equilibrium, $F$ takes the form $F_0=-m g$ in order to compensate gravity, where $m$ refers to the mass of the rigid body and $g$ to the gravity vector. The rotation matrix describing the orientation of the body-fixed frame with respect to the inertial frame (that is oriented such that at equilibrium the inertial and body-fixed frames agree) is parametrized with the xyz-Euler angles $\varphi_1,\varphi_2$, and $\varphi_3$.\footnote{Note that a different parametrization in terms of quaternions or a rotation vector leads to the same linearized dynamics.} Linearizing the dynamics about the ``upright" equilibrium yields
\begin{align}
\ddot{\varphi}_1(t)&=\pi_1^2 \varphi_1(t) + \pi_1^2 ~\frac{\V[F]{I}_2(t)}{m g_0},\\
\ddot{\varphi}_2(t)&=\pi_2^2 \varphi_2(t) - \pi_2^2 ~\frac{\V[F]{I}_1(t)}{m g_0},
\end{align}
with $\pi_1^2=l m g_0/I_1$ and $\pi_2^2=l m g_0/I_2$, and where $\V[\omega]{B}=(\dot{\varphi}_1,\dot{\varphi}_2,\dot{\varphi}_3)$, $g_0=9.81$m/s$^2$, and $\V[F]{I}_1$ and $\V[F]{I}_2$ represent the first two components of the force $F$ expressed in the inertial frame.

This corresponds to \eqref{eq:sys1} and \eqref{eq:sys2} if $\V[F]{I}_1(t)$ and $\V[F]{I}_2(t)$ happen to be linearly dependent. We can then conclude that in order to maximize controllability (in the sense discussed previously), the inertia $I_1$ and $I_2$ should be chosen such that 
\begin{equation}
\frac{I_1}{I_2}=\delta_\text{s}^2=(1+\sqrt{2})^2.
\end{equation}

More concrete examples of such cases cases include rocket-type systems or balancing robots (e.g. \cite{rezero}), where the actuation is severely limited (for example due to failures). This also comprises the inverted-pendulum-on-a-cart system with two pendula of different lengths in parallel, \cite{barton}, and the three-dimensional reaction-wheel inverted pendulum, \cite{cubli}.

\section{Conclusion}
This note investigated the controllability of a certain type of inverted pendulum systems. It was shown that controllability (in the sense described earlier) is maximized when the ratio between the two time-constants equals the silver ratio.

We leave it to the reader to think about a geometric interpretation of the result and to philosophize about its deeper significance (if there is one at all).

\section*{Acknowledgment}
The author would like to thank the Branco Weiss Fellowship - Society in Science, administrated by ETH Zurich for the generous support.

\bibliographystyle{IEEEtran}
\bibliography{literature}

\end{document}